\documentclass[prb,a4paper,showpacs,superscriptaddress]{revtex4}
\usepackage{amsmath} 
\usepackage{amsfonts}
\usepackage{amssymb}
\usepackage{bm} 
\usepackage{graphicx}
\newcommand{\nambu}{\breve}
\newcommand{\unit}{\hat}
\newcommand{\spin}{\underline}
\newcommand{\imagu}{{\rm i}}
\newcommand{\epm}{\epsilon_m}

\begin{document}
\title{Model of Inhomogeneous Impurity Distribution in Fermi Superfluids}
\author{R. H\"anninen}
\affiliation{Low Temperature Laboratory, Helsinki University of
Technology, P.O.Box 2200, FIN-02015 HUT, Finland } 
\author{E. V. Thuneberg}
\affiliation{Low Temperature Laboratory, Helsinki University of
Technology, P.O.Box 2200, FIN-02015 HUT, Finland } 
\affiliation{Department of Physical Sciences,
P.O.Box 3000, FIN-90014 University of Oulu, Finland}

\date{\today}

\begin{abstract}
The standard treatment  of impurities in metals assumes a homogeneous
distribution of impurities. In this paper we study distributions that
are inhomogeneous.   We discuss in detail the ``isotropic
inhomogeneous scattering model'' which  takes into account the
spatially varying scattering on the scale of the superfluid coherence
length. On a large scale the model  reduces to a homogeneous medium
with renormalized parameter values. We apply the model to superfluid
$^3$He, where  porous  aerogel acts as the impurity. We calculate the
transition temperature $T_{\rm c}$, the order parameter, and the
superfluid density. Both A- and B-like phases are considered. Two
different types of behavior are identified for the temperature
dependence of the order parameter.  We compare the calculations with
experiments on $^3$He in aerogel. We find that most of the differences
between experiments and the homogeneous theory can be explained by the
inhomogeneous model. All our calculations are based on the
quasiclassical theory of Fermi liquids. The parameters of this theory
for superfluid $^3$He in aerogel are discussed. 
\end{abstract}

\maketitle

\section{Introduction}\label{s.int}

The standard treatment of impurities in a metal is based on averaging
over the locations of the impurities.\cite{AGD} This means that the
probability of a quasiparticle 
being scattered is independent of its location. To be definite, we call this
the {\em homogeneous scattering model} (HSM). The purpose of this paper is
to study the case where the impurity distribution varies in space. We
define a model of inhomogeneous scattering, which is as simple as possible
and represents a medium that is uniform and isotropic on a large scale.
We call it the {\em isotropic inhomogeneous 
scattering model} (IISM).\cite{ErkkiModels}
The model is motivated by experiments on superfluid $^3$He in
a porous aerogel, and all our numerical computations concentrate on this
case. However, the model is independent of the pairing symmetry and
therefore can equally be applied, for
example, to $s$- or $d$-wave superconductors.  

Liquid $^3$He is an ideal example of unconventional superfluid because it is
naturally pure, it has a spherical Fermi surface, and its $p$-wave
pairing state is well understood. Therefore it is of interest to study
impurity effects in this superfluid. The addition of impurities to helium
can be done by using porous aerogel so that typically 98\%  of the volume is
occupied by
$^3$He. It was found experimentally that the superfluid transition
temperature in aerogel is reduced but remains sharp.\cite{Porto95} Also other
superfluid properties such as the superfluid density and the NMR shifts were
found to be modified by aerogel. Many experiments studying this system have
been made during the last seven years.\cite{halprev}

The HSM is an attractive model for $^3$He in aerogel because of its
simplicity. Essentially all theoretical calculations for this system
are based on it. \cite{ErkkiModels,baramidze,M,ErkkiGL,PhysicaB,raineraero,BK2,
higashitani,SharmaSauls,mineevkrotkov,Baramidze2,SharmaLT23,yamamoto}
However, already the first comparisons to experiment showed that the HSM is
insufficient quantitatively. In particular, it was found that the order
parameter is more suppressed than the superfluid transition temperature
$T_{\rm c}$. A natural explanation for this comes from the fact that the
scattering in aerogel is not homogeneous, as has already been demonstrated
using the IISM.\cite{ErkkiModels}  Similar results have been reached in
calculations on unconventional superconductors.
\cite{FKBS}

Preliminary results of the IISM have been reported in Refs.\
\onlinecite{ErkkiModels} and \onlinecite{LT22}. In  this paper we
present the IISM in detail. Our studies are based on the 
quasiclassical theory of Fermi liquids. We discuss the assumptions of this
theory (Sec.\ \ref{s.quasiclas}), and how these are satisfied  for
$^3$He in aerogel (Sec.\ \ref{s.aerogel}).  The inhomogeneous scattering model
is introduced in Sec.\ \ref{s.iism}. For the case of $p$-wave pairing we
introduce the order parameters of A- and B-type phases (Sec.\
\ref{s.op}). We calculate several quantities including the critical
temperature, order parameter, and superfluid density (Sec.\
\ref{s.res}). The results are compared with experiments on $^3$He in
aerogel (Sec.\ \ref{s.exp}).  The equations of quasiclassical theory
and the details of calculations are discussed in the Appendix.  

\section{Quasiclassical theory}\label{s.quasiclas}

On a microscopic scale a pure system is described in terms of particles
(conduction electrons or $^3$He atoms) and their interaction. Because the
interactions are strong, this leads to a
complicated many-body problem. 

The characteristic length in superconductivity or superfluidity is the
coherence length. We define this quantity as
\begin{equation}
\xi_0=\frac{\hbar v_{\rm F}}{2\pi k_{\rm B}T_{\rm c0}},
\label{e.cohlength}\end{equation}
where $v_{\rm F}$ is the Fermi velocity. To be precise, we have used the
superfluid transition temperature $T_{\rm c0}$ of a pure system.
The coherence length is typically much larger than the Fermi wave length
$\lambda_{\rm F} = 2\pi/k_{\rm F}$. In $^3$He $\xi_0$ depends on pressure
and changes from 16 nm at the solidification pressure to 77 nm at the vapor
pressure whereas $\lambda_{\rm F} \approx 0.7$ nm. 

The theory that is designed to work on the scale $\xi_0\gg\lambda_{\rm
F}$  is the quasiclassical theory.\cite{Serene} It treats
the system as a dilute gas of weakly interacting quasiparticles. In
quasiclassical theory all the  many-body physics that takes place on the
microscopic scale $\lambda_{\rm F}$ is eliminated. It only appears
through phenomenological parameters like the Fermi-surface, Landau
Fermi-liquid parameters and transition temperature $T_{\rm c0}$. 

Let us consider any external objects in the system. These objects are
characterized by a strong potential, on the order of the Fermi energy. As a
consequence the state of the system is modified in the vicinity of the
object. A theoretical analysis of these atomic scale changes is again
difficult because of the strong interactions between particles. 

In the quasiclassical theory the effect of external objects is
twofold.\cite{BR} First, the phenomenological parameters 
discussed above are changed. These parameters are determined by processes on
the Fermi-energy scale and are therefore of short range. Assuming that the
surface area of the external objects times the atomic length scale
($\lambda_{\rm F}$) is a small fraction of the total volume,
this effect is small and is neglected in the following. Second, the
objects affect the low energy processes (energy $\sim k_BT_{\rm c0}$)
directly via a scattering of quasiparticles. The range of this effect is long,
on the order of the coherence length [Eq. (\ref{e.cohlength})]. Therefore 
it leads to substantial modification of the superfluid properties.

An important length characterizing the scattering is the mean free path
$\ell$. The simplest case is to consider the limit  $\lambda_{\rm
F}/\ell\rightarrow 0$. This is the quasiclassical limit where the Fermi wave
length effectively disappears from the theory. First order corrections
in $\lambda_{\rm F}/\ell$ lead to effects like weak localization,
which are neglected here. 
In quasiclassical theory, the scattering is represented by a collection of
``scattering centers.'' The main assumption is that the quantum interference
between two scattering centers is neglected. Technically this can be achieved
considering an ensemble average where the locations ${\bf r}_i$ of the
scattering centers are uncertain by a distance on the order of $\lambda_{\rm
F}$ or more. The size of one scattering center is limited by the condition
that it has to be localized on the scale of
$\xi_0$. The scattering properties of a center can be parametrized by
scattering phase shifts $\delta^{(l)}$, which are taken at the Fermi energy
in the normal state. (For simplicity we label the different partial waves by
a single index $l$, but there is no need to restrict to spherically symmetric
scattering centers. The scattering could also be spin-dependent, but it is
also neglected here for simplicity.) Thus a complete description of the
scattering needed in the quasiclassical theory consists of distribution
functions $n_i({\bf r})$ and scattering phase shifts $\delta_i^{(l)}$
of the scattering centers $i=1,2,\ldots$. 

Within each scattering center an exact quantum treatment is allowed in
principle. However, because of uncertainty about the microscopic
processes at surfaces, the phase shifts $\delta_i^{(l)}$ cannot be
calculated from first principles. Instead, one has to use some models,
for example hard spheres.\cite{Merzbacher} In some quantities the
phase shifts only appear in certain combinations, for example, the
transport cross section 
\begin{equation}
\sigma=\frac{4\pi}{k_{\rm F}^2}\sum_{l=0}^\infty(l+1)
\sin^2(\delta^{(l+1)}-\delta^{(l)}).  
\label{e.sigmatr}\end{equation}

Let us try to clarify some consequences of the assumptions made above. 
The energy one usually is trying to calculate is on the order of $f_{\rm
cond}\xi_0^3$. Here $f_{\rm cond}\sim k_{\rm B}T_{\rm c}/\lambda_{\rm
F}^2\xi_0$ is the superfluid condensation energy per volume, and a typical
volume $\sim\xi_0^3$. Above the individual scattering centers were
required to be small, $\sigma\ll\xi_0^2$. This implies that the energy
$f_{\rm cond}\sigma\xi_0$ associated with a single impurity\cite{TKR}
is small in comparison. The typical number of impurities in volume
$\xi_0^3$ is $N\sim\xi_0^2/\sigma$, which for random impurities
implies a fluctuation $\delta N\sim\sqrt{N}$. The corresponding
fluctuation in energy is by factor $\sqrt{\sigma}/\xi_0$ smaller than
$f_{\rm cond}\xi_0^3$.  This has to be neglected since there are other
neglected contributions that are on the same order of magnitude. Thus 
impurity averaging in the quasiclassical approximation implies  a scattering
medium where the fluctuations in the impurity density are neglected.

\section{Aerogel}\label{s.aerogel}

The structure of aerogel, as relevant for $^3$He experiments, is  discussed
in Ref.\ \onlinecite{Porto99}. Here we repeat some main points. Aerogel
consists of small SiO$_2$ particles of diameter $\sim 3$ nm, which are
coalesced together to form a self-supporting structure. Experiments with
$^3$He typically use aerogels with open volume fraction 98\% or more.
According to small-angle x-ray scattering measurements, there is a
``fractal'' range in the particle cluster up to a ``correlation length''
$\xi_a\sim 100$ nm. Above this scale the structure looks homogeneous. 
Computer simulations give a picture of widely spaced aerogel strands
implying  a long mean free path $> 100$ nm.

It seems reasonable that the quasiclassical description above can be applied
to liquid $^3$He in aerogel. The atomic layer on the  SiO$_2$ surfaces
occupies only one per cent of the volume and can be neglected for many
purposes. (Magnetic properties make an exception because the
susceptibility in this layer is much larger than in pure
liquid.\cite{Sprague95})  Since a major part of the liquid is within a
coherence length from SiO$_2$,\cite{Porto99} the scattering effect leads to
a substantial modification of the superfluid properties.  It also seems that
the scattering from aerogel can be represented by incoherent scattering
centers whose size is small compared to
$\xi_0$ because random variations on the order of $\lambda_{\rm F}$ are
likely to develop already at much smaller distances. 

The smallest reasonable choice for a scattering center is a single SiO$_2$
particle (diameter $2R_a\sim 3$ nm). This is large in comparison to
$\lambda_{\rm F}$: $k_{\rm F}R_a\sim 10$. According to hard sphere
phase shifts this means that only $1\%$ of the scattering takes place
in the s-wave channel, and $99\%$ is left to higher partial waves in
Eq.\ (\ref{e.sigmatr}). This dominant contribution of higher partial
waves has several important consequences.

Firstly, the phase shifts $\delta_i^{(l)}$ {\em are random numbers}.
This is because even the phase shifts (modulo $\pi$) of a hard sphere with a
fixed $R_a$ are pseudorandom numbers for $k_{\rm F}R_a\gg 1$. Adding
to this the varying particle size, the surface roughness and varying
orientations of touching neighbor particles, it is simply impossible
that the result would be anything else but random. (The randomness, of
course, is valid only for partial waves $l$ that contribute
essentially to scattering, i.e., for $l<k_{\rm F}R_a$.)

A consequence of the random phase shifts is that only the number of
scattering phase shifts, or equivalently, the cross section $\sigma$
[Eq. (\ref{e.sigmatr})] is important in describing a scattering center.  Thus
a sufficient description of the scattering is obtained by specifying only
$\sigma_i$ and
$n_i({\bf r})$. 
 
A second important consequence of the large particle size is that both
$\sigma_i$ and $n_i({\bf r})$ are independent of  pressure. This is crucial
for comparison with experiments, because when fitting is needed, it can be
done at one pressure only, and the predictions of the model get fixed at all
pressures. We argue as follows. It is reasonable to assume that the aerogel
is independent of the hydrostatic pressure, implying that $n_i({\bf r})$ is
also. The Fermi wave vector changes by 10\% over the pressure range from
the vapor pressure to the solidification pressure. This could induce some
pressure dependence in the cross section $\sigma_i$. For example, hard-sphere
$\sigma$ depends essentially on $k_{\rm F}$ in the region $k_{\rm F}R_a\sim
1$.\cite{Salomaa} However, in the physical region
$k_{\rm F}R_a\gtrsim 10$ the  dependence of $\sigma$ on
$k_{\rm F}$ is very weak. Thus we conclude that both 
$\sigma_i$ and $n_i({\bf r})$ are independent of pressure. 

The large $k_{\rm F}R_a$ is potentially bad news for theory because
calculations that take higher partial waves into account are very
complicated, see Ref.\ \onlinecite{PhysicaB}. The promising conclusion
of these calculations is that, at least in some cases, the results
including higher partial waves are almost identical to those including
only s-wave {\em under the following conditions}: (i) one uses the
same transport mean free path $\ell$ and (ii) one uses either random
phase shift $\delta^{(0)}$ or fixed $\sin^2\delta^{(0)} \approx
\frac{1}{2}$ in the s-wave calculation. Here we assume that this
correspondence holds more generally. Thus we calculate only s-wave and
present results for $\sin^2\delta^{(0)} = \frac{1}{2}$. 
Finally, the randomness of the phase shifts also simplifies the numerical
calculations since it implies that some components of the propagator vanish
(see the Appendix).  

The use of several impurity species is important in studying 
anisotropic scattering where the preferred direction varies in space.
In this paper we limit to isotropic scattering. In order to simplify the
notation we select all scattering centers to have equal cross section, so
that the only scattering parameters are
$\sigma$ and the total impurity density $n({\bf r})$.

\section{Setting the model}\label{s.iism}

The simplest possible impurity profile $n({\bf r})$ is a constant. It
implies a location independent (transport) mean free path 
$\ell=(n\sigma)^{-1}$.
This {\em homogenous scattering model} (HSM) has successfully
been used to model the impurities in
superconductors.\cite{AGD,general} For  
impure $p$-wave superfluid ($^3$He) it has been used to calculate
the critical temperature,\cite{larkin} the order parameter and superfluid
density\cite{ErkkiModels,PhysicaB,raineraero,higashitani}, 
properties near the superfluid transition,\cite{M,ErkkiGL}
properties in magnetic
field,\cite{baramidze,BK2,SharmaSauls,mineevkrotkov,yamamoto}
density of states,\cite{SharmaSauls,mineevkrotkov}
thermal conductivity\cite{SharmaLT23}
and estimate strong-coupling effects.\cite{Baramidze2} 

The predictions of the HSM are compared to experiments in Refs.\
\onlinecite {ErkkiModels,ErkkiGL,raineraero,higashitani,SharmaSauls,
mineevkrotkov,SharmaLT23,Porto99,Hook} and \onlinecite{Lawes}, see also
below. Compared to pure $^3$He, the HSM gives the right tendency and can
work even quantitatively in some cases, but there can be differences up to a
factor of 5 in the suppression factors (see below). In some
papers the experiments are compared with data that is calculated in the
unitary limit of s-wave scattering, which seems to give better agreement
than other phase shifts. As explained above, we believe this is misguided,
and random or intermediate phase shifts should be used instead. The observed
differences can more naturally be explained by inhomogeneous scattering as
will be shown below, at least for some quantities.

Real aerogel has voids where scattering is negligible. This can be
modelled by an impurity density $n({\bf r})$ which depends on the
location ${\bf r}$.  It is in principle possible to use a realistic
$n({\bf r})$ for aerogel. This has the drawback that the computational
volume should be large in order to get a representative sample,
and this implies heavy numerical effort. Here we prefer the opposite
limit of a simple model $n({\bf r})$. The simplest possibility would be
a plane-wave variation $n({\bf r})=n_0+n_1\cos({\bf q}\cdot{\bf r})$.  A
stronger version of this would be equally spaced scattering planes. In the
limit of very strong scattering in the planes this leads to isolated
slabs.\cite{ErkkiModels}
The problem with all these models is that they are anisotropic. For
example, the superfluid density would depend on the direction of the
superfluid velocity ${\bf v}_s$. In comparison to experiments one should use
some average over the directions, but this neglects the process how the
averaging really takes place by a nonuniform current  distribution.

The purpose of the {\em isotropic inhomogeneous scattering model}
(IISM) (Ref. \onlinecite{ErkkiModels}) is to incorporate a non-constant $n({\bf r})$
with spherical symmetry. We take a spherical volume of radius $R$ and
use an impurity density $n(r)$ that depends only on the radial
coordinate $r$. We assume that these spheres fill all the space. This
last point is not strictly possible, but represents an approximation
that is similar to using spherical approximation for a Wigner-Seitz
unit cell.\cite{AM}  The calculation in a single unit cell can
represent states where the superfluid order parameter $\tensor A({\bf
r})$ has the Bloch form
\begin{equation}
\tensor A({\bf r})=\tensor A^{(0)}({\bf r})\exp(i{\bf q}\cdot{\bf r}),
\label{e.bloch}\end{equation}
where $\tensor A^{(0)}({\bf r})$ is a strictly periodic order parameter.
In the present case the wave vector ${\bf q}$ is the imposed phase gradient
that is related to the superfluid velocity ${\bf v}_{\rm s} =
(\hbar/2m){\bf q}$ defined on a scale larger than $R$.

\begin{figure}[tb]
\begin{center}
\includegraphics[width=4cm]{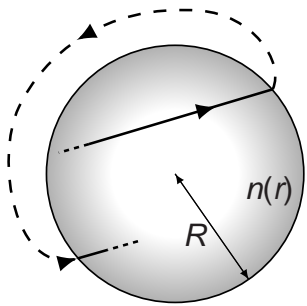}
\caption{The sphere used in the IISM to approximate a unit cell of a
periodic lattice. The drawing illustrates the boundary condition that a
quasiparticle exiting the sphere is returned back at the diametrically
opposite point. The shading depicts the spherically symmetric impurity
density $n(r)$. }
\label{f.sphereiism}
\end{center}
\end{figure}

At the surface of the IISM sphere we impose the boundary condition that
an exiting quasiparticle is effectively returned to the sphere at the 
diametrically opposite point, see Fig.\
\ref{f.sphereiism}. In the case of current-carrying states,
there has to be a phase shift which corresponds to the phase factor in
Eq.\ (\ref{e.bloch}). Otherwise the state of the quasiparticle (momentum,
spin)  is unchanged. The boundary condition can be expressed mathematically
for the Green's function as 
\begin{eqnarray}
\breve{g}({\hat {\bf k}}, R\hat{\bf r}, \epsilon_m) =\exp\left(
{\rm i}{\bf q} \cdot R\hat{\bf r}\, \breve{\tau}_3 \right)
\breve{g}({\hat {\bf k}}, -R\hat{\bf r}, \epsilon_m) \exp\left(
-{\rm i}{\bf q} \cdot R\hat{\bf r} \,\breve{\tau}_3 \right) ,
\label{e.boundary}
\end{eqnarray}   
see the Appendix for notation.
The model reduces to the HSM in the limit that $n(r)$ is independent of $r$.
 
Because of the spherical approximation, it is worth while to properly
define all large-scale ($\gg R$) quantities. We define macroscopic
quantities  as averages of the corresponding local quantities over the
sphere, 
\begin{equation}
\langle A\rangle\equiv\frac{3}{4\pi R^3}\int_{r<R} d^3 r
A({\bf r}).
\end{equation}
In particular, corresponding to the local mass current density of the
superfluid, ${\bf j}_s$, we have the macroscopic $\langle{\bf
j}_s\rangle$. The superfluid density $\tensor \rho_s$ is then defined
by $\langle{\bf j}_s\rangle=\tensor \rho_s{\bf v}_s+O(v_s^2)$ with
${\bf v}_{\rm s} = (\hbar/2m){\bf q}$ as defined above.
The average mean free path $\ell_{\rm ave}$ is defined by $\ell_{\rm
ave}^{-1} = \sigma \langle n(r) \rangle$.

The parameters specifying the scattering are the radius of the sphere
$R$, the average mean free path $\ell_{\rm ave}$, and the shape of the
impurity density $n(r)$.  For $n(r)$ we use two different analytic forms 
\begin{eqnarray}
n(r) &=& c\left[ \left(\frac{r}{R}\right)^j -
\frac{j}{j+2}\left(\frac{r}{R}\right)^{j+2} \right],
\label{e.idens1} \\
n(r) &=& c' \left[ \cos^j\left(\frac{\pi r}{2 R}\right) + b \right], 
 j \ge 2 .
\label{e.idens2}
\end{eqnarray}
with parameters $j$ and $b$. Here the prefactors $c$ and $c'$ are
determined by the parameters $\ell_{\rm ave}$ and $\sigma$ ($c$,
$c'>0$). The functions in Eqs.\ (\ref{e.idens1}) and (\ref{e.idens2})
are shown in the inset of Fig.\ \ref{f.gapesim}. We call them {\em
void} and {\em cluster} profiles, respectively, because the former has
strongest scattering at the boundary and the latter at the center of
the sphere. Both profiles have zero derivative at $r=R$ in order to
have a smooth impurity density everywhere. 

An attractive feature of the IISM is that  the symmetries of a homogeneous
system are preserved on a large scale $\gg R$. Thus one can apply
phenomenological large-scale theories such as Ginzburg-Landau and
hydrodynamic theories as for a homogeneous
medium.\cite{Volovik,ErkkiGL,fomin}  The only change is that the parameters
of these theories are modified by the inhomogeneity. Some of these
parameters are calculated below.

\section{Order parameter}\label{s.op}

We apply the IISM for superfluid $^3$He.
The main assumption in addition to those already mentioned is the
use of the weak-coupling approximation. The dipole-dipole interaction is
neglected because it is unimportant on the scale of a few $\xi_0$, which we
study here. The equations and some details of the numerical
implementation are discussed in the Appendix.

Even with all simplifications, the computational effort in the IISM is quite
substantial. For example, the computer code has five nested loops in
addition to the one needed for the iteration of the order parameter, and
there are several stages of initialization, interpolation, and data
collection. 

\begin{figure}[tb]
\begin{center}
\includegraphics[width=8cm]{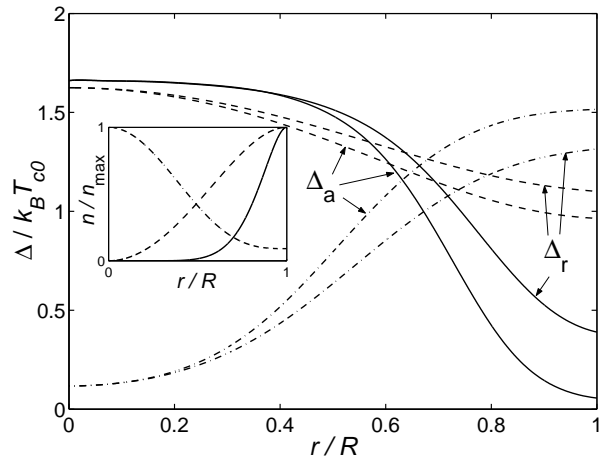}
\caption{The B-phase order parameter [Eq. (\ref{ordparBvszero})]
in the IISM for ${\bf v}_s=0$ at the temperature $0.5 T_{\rm
c0}$. The solid lines are for {\em steep void} impurity profile 
[Eq.\ (\ref{e.idens1}) with $j = 8$], $R = 2\ell_{\rm ave}$, and
$\xi_0/\ell_{\rm ave}=0.2$. The dashed lines are for {\em gentle void}
profile [Eq.\ (\ref{e.idens1}) with $j = 2$], $R = \ell_{\rm ave}$, and
$\xi_0/\ell_{\rm ave}=0.1$. The dash-dotted lines are for {\em cluster}
profile [Eq.\ (\ref{e.idens2}) with $j = 3$, $b = 0.1$], $R = \ell_{\rm
ave}$, and $\xi_0/\ell_{\rm ave}=0.1$. The inset shows the
corresponding impurity profiles.}
\label{f.gapesim}
\end{center}
\end{figure}

Based on calculations with the HSM, no new superfluid phases
of $^3$He are expected in the presence of scattering.\cite{ErkkiModels}
However, the order parameters of the A and B phases are
modified by inhomogeneous scattering. The general forms can be deduced
using symmetry arguments. The B phase order parameter for
${\bf v}_s=0$ has the form
\begin{eqnarray}
\tensor{A}(r,\phi,\theta) =e^{{\rm i} \chi}\tensor{R}\left[ \Delta_{r}(r)
\hat{\bf r} \hat{\bf r} + \Delta_{a}(r) (\hat{\boldsymbol{\theta}}
\hat{\boldsymbol{\theta}}  + \hat{\boldsymbol{\phi}}
\hat{\boldsymbol{\phi}})\right], 
\label{ordparBvszero}
\end{eqnarray}
where spherical coordinates ($r,\phi,\theta$) are used. The phase
$\chi$ and the rotation matrix $\tensor{R}$ are
arbitrary constants. The calculation determines the real-valued    
radial $\Delta_{r}(r)$ and angular $\Delta_{a}(r)$ functions. These functions
are shown in  Fig.\ \ref{f.gapesim} for three different scattering profiles
$n(r)$. 
We see that the order parameter is inhomogeneous, and is mostly
suppressed in regions where the scattering is strong. For the void profile
[Eq. (\ref{e.idens1})] the order parameter components have maxima
$\Delta_{\rm max}=\Delta_{r}(0)=\Delta_{a}(0)$ at the center and minima
at the surface. The absolute minimum $\Delta_{\rm min}$
is given by $\Delta_{a}(R)$.

We define two different averages of the spatially varying order
parameter. The usual average is given by
\begin{equation}
\Delta_{\rm
ave}^2=\frac{1}{3}\langle\Delta_{r}^2+2\Delta_{a}^2\rangle. 
\label{e.deltaave}\end{equation}
An ``NMR average'' is defined by
\begin{equation}
\Delta_{\rm NMR}^2
=\frac{1}{15}\langle
2\Delta_{r}^2+6\Delta_{r}\Delta_{a}+7\Delta_{a}^2\rangle.
\label{e.deltanmr}\end{equation}
This form can be justified as follows. The frequency shifts of nuclear
magnetic resonance in superfluid $^3$He are determined by the 
dipole-dipole interaction energy \cite{Leggett}
\begin{equation}
f_{\rm d}=g_{\rm d}[\vert\mathop{\rm Tr}\tensor A\vert^2
+\mathop{\rm Tr}(\tensor A^*\tensor A)].
\label{e.fdbulk}\end{equation}
The coefficient $g_{\rm d}$ is a phenomenological parameter that depends on
a cut-off energy, which is on the order of the Fermi energy.\cite{btex}
According to the principles of Sec.\ \ref{s.quasiclas} we assume it is not
changed by the impurity. We calculate the dipole-dipole energy
[Eq. (\ref{e.fdbulk})] for the B-phase order parameter (\ref{ordparBvszero}) and
find 
\begin{equation}
f_{\rm d}=4g_{\rm d}\Delta_{\rm
NMR}^2\cos\vartheta(1+2\cos\vartheta),
\label{e.ddb}\end{equation}
where $\vartheta$ is the rotation angle of $\tensor{R}$. This has
exactly the same form as in pure $^3$He-B, the only change being 
that the energy gap $\Delta$ is replace by $\Delta_{\rm NMR}$. Thus
the NMR properties within IISM are the same as in bulk liquid except
for this renormalization. 

We point out that the order parameter in Eqs.\
(\ref{ordparBvszero})--(\ref{e.ordparB}) is
defined using the off-diagonal part of the mean-field self energy
(\ref{e.purese}). It should be noted that the order parameter is not
simply related to the energy gap in the excitation spectrum. Only in
pure, homogeneous superfluid the energy gap equals
$\Delta_{\rm ave}=\Delta_{\rm NMR}$.  The
excitation spectrum in the HSM has been studied in Refs.\
\onlinecite{SharmaSauls} and
\onlinecite{mineevkrotkov}, but there are no studies yet for the IISM. 

The A-phase order parameter is more complicated since it is
anisotropic. We select the cylindrical coordinates ($\rho,\phi,z$)
so that $z$ is along the anisotropy axis $\hat{\bf l}$. The order
parameter for ${\bf v}_s=0$ can be written as 
\begin{eqnarray}
\tensor{A}(\rho,\phi,z) = e^{{\rm i} \phi} \hat{\bf d}\left[ 
\Delta_{\rho}(\rho,z) \hat{\boldsymbol{\rho}} +
{\rm i} \Delta_{\phi}(\rho,z) \hat{\boldsymbol{\phi}} +
\Delta_{z}(\rho,z) \hat{\bf z} \right], 
\label{e.ordparA}
\end{eqnarray}
where $\hat{\bf d}$ is an arbitrary constant unit vector. The functions
$\Delta_{i}$ are real but now they depend on two coordinates $\rho$
and $z$. They satisfy symmetry relations
$\Delta_{\rho}(\rho,-z) = \Delta_{\rho}(\rho,z)$,
$\Delta_{\phi}(\rho,-z) = \Delta_{\phi}(\rho,z)$ and
$\Delta_{z}(\rho,-z) = - \Delta_{z}(\rho,z)$. For the A phase we define  
\begin{eqnarray}
\Delta_{\rm
ave}^2&=&\frac{1}{2}\langle
\Delta_{\rho}^2+\Delta_{\phi}^2+\Delta_{z}^2\rangle
\label{e.deltaaveA}\\
\Delta_{\rm
NMR}^2&=&\frac{1}{2}\langle
\Delta_{\rho}^2+\Delta_{\phi}^2-2\Delta_{z}^2\rangle.
\label{e.deltanmrA}\end{eqnarray}
In pure homogeneous superfluid $\Delta_{\rm
ave}=\Delta_{\rm NMR}$ equals the maximum energy gap in the A phase.
The dipole-dipole energy [Eq. (\ref{e.fdbulk})] in the A phase
(\ref{e.ordparA}) is given by
\begin{equation}
f_{\rm d}=-2g_{\rm d}\Delta_{\rm NMR}^2(\hat{\bf d}\cdot\hat{\bf
l})^2,
\label{e.dda}\end{equation}
and the justification for $\Delta_{\rm NMR}$ (\ref{e.deltanmrA}) is
completely analogous to the case of the B phase.  

\begin{figure}[tb]
\begin{center}
\includegraphics[width=6cm]{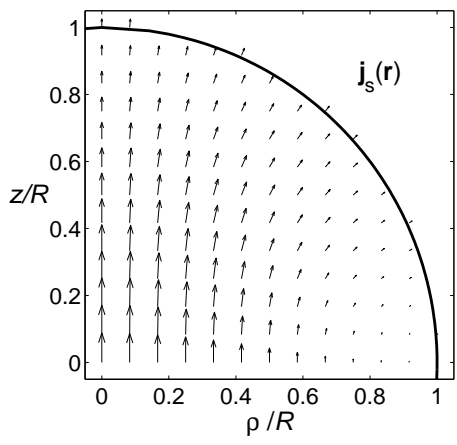}
\caption{The local superflow ${\bf j}_{\rm s}(\rho,\phi,z)$ in
the IISM unit cell.  The parameters are the steep void impurity  profile
[Eq. (\ref{e.idens1}) with $j=8$], $R = \ell_{\rm ave} = 5\xi_0$,
$T=0.6T_{\rm c0}$, $F_1^{\rm s}=10$ (pressure 15.6 bar), and $q\xi_0 =
0.001$.   
}      
\label{f.curresim}
\end{center}
\end{figure}

In the case of a finite superfluid velocity (taken to be in the
$z$ direction) we limit our calculations to the B phase where the
order parameter takes the form 
\begin{eqnarray}
\tensor{A}(\rho,\phi,z) =e^{{\rm i} \chi}\tensor{R}\left[
\Delta_{\rho\rho} \hat{\boldsymbol{\rho}}\hat{\boldsymbol{\rho}} +
\Delta_{\phi\phi} \hat{\boldsymbol{\phi}}\hat{\boldsymbol{\phi}} +
\Delta_{zz} \hat{\bf z}\hat{\bf z} + 
\Delta_{\rho z} \hat{\boldsymbol{\rho}} \hat{\bf z} +
\Delta_{z \rho} \hat{\bf z} \hat{\boldsymbol{\rho}}\right] . 
\label{e.ordparB}
\end{eqnarray}
Now $\Delta_{\mu i}$ are complex and depend on $\rho$ and $z$. They
satisfy symmetry relations  $\Delta_{\rho z}(\rho,-z) = -\Delta_{\rho
z}^{*}(\rho,z)$, $\Delta_{z\rho}(\rho,-z) = -
\Delta_{z\rho}^{*}(\rho,z)$ and $\Delta_{ii}(\rho,-z)  =
\Delta_{ii}^{*}(\rho,z)$, where $i = \rho,\phi,z$. The mass
supercurrent ${\bf j}_{\rm s}({\bf r})$ has non-zero $\rho$ and
$z$-components inside the sphere. One case is illustrated in Fig.\
\ref{f.curresim}. 
It can be seen that the current density is smallest in regions of
strong scattering. Naturally, the current is conserved, 
${\boldsymbol{\nabla}}\cdot{\bf j}_{\rm
s}({\bf r})=0$. In the void profile, the current has to go through the
scattering regions at the cell boundary, whereas in the cluster
profile (not shown) the transport current can flow past the scattering
region situated at the center.  Regardless of the profile, the averaged
current is parallel to ${\bf v}_s$ and independent of the
direction. Thus the superfluid density tensor $\tensor \rho_s$ reduces
to a scalar $\rho_s$.

\section{Results}\label{s.res}

\begin{figure}[tb]
\begin{center}
\includegraphics[width=8cm]{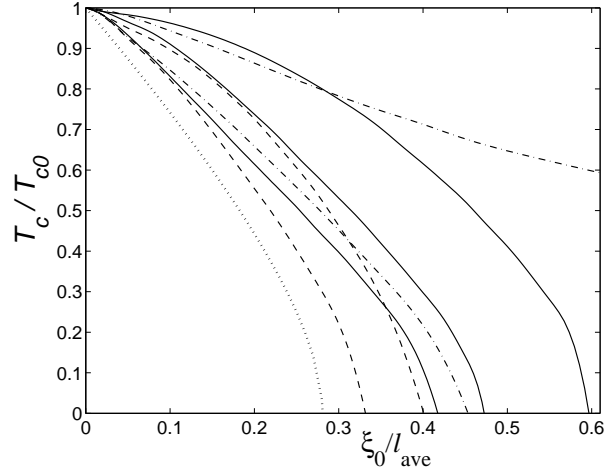}
\caption{The transition temperature as a function of
average scattering rate $\xi_0/\ell_{\rm ave}$. The dotted line is
from the HSM. Solid lines are for the steep void impurity profile
[Eq.\ (\ref{e.idens1}) with $j=8$], and correspond to $R/\ell_{\rm
ave} =\frac{1}{2}$, 1, and 2, from left to right. The dashed lines are
for the gentle void profile [Eq.\ (\ref{e.idens1}) with $j=2$] and
from left to right $R/\ell_{\rm ave} = 1$ and 2. The dash-dotted lines
are for cluster impurity profile  Eq.~(\ref{e.idens2}) with $j=3$ and
$R=\ell_{\rm ave}$. The lower curve is with $b = 0.1$ and the upper
one with $b = 0$.}  
\label{f.tcsiism}
\end{center}
\end{figure}

We start by  studying the critical temperature $T_{\rm c}$. We calculate
the quantity $T_{\rm c}/T_{\rm c0}$,  the critical temperature relative to
the critical temperature in the absence of scattering. In the HSM this
depends only on the parameter $\xi_0/\ell$. The dependence turns out to be
the same as calculated by Abrikosov and Gorkov for $s$-wave superconductors in
the presence of magnetic impurities.\cite{AG} This result was generalized to
the nonmagnetic $p$-wave case in Ref.\ \onlinecite{larkin}. In the
IISM $T_{\rm c}/T_{\rm c0}$ depends only on the ratio
$\xi_0/\ell_{\rm ave}$ and the impurity profile
$n(r)$. In particular, it is independent of the phase (A or B) and the phase
shifts $\delta^{(l)}$. The relative $T_{\rm c}$ is plotted  as a function of
$\xi_0/\ell_{\rm ave}$ in Fig.\ \ref{f.tcsiism} for different
scattering profiles.
One sees that in all cases the HSM gives the lowest
$T_{\rm c}$. This is natural since the inhomogeneity implies existence of
regions where the scattering is less than the average, and in these
regions the order parameter nucleates at a higher temperature. For example,
for the steep void profile with $R/\ell_{\rm ave} =2$ we find $T_{\rm
c}=0.8T_{\rm c0}$ at the average impurity where superfluidity in the HSM
is completely suppressed ($\xi_0/\ell_{\rm ave}=0.28$). Generally one
concludes that the larger the inhomogeneity (in amplitude and in
length scale), the larger $T_{\rm c}$ is obtained at a given
$\xi_0/\ell_{\rm ave}$. One notices that the cluster profile with no
background scattering ($b = 0$) differs qualitatively from the other
curves in  Fig.\ \ref{f.tcsiism}. This profile is exceptional  because
it has long quasiparticle trajectories where the scattering is negligible.   

\begin{figure}[tb]
\begin{center}
\includegraphics[width=8.5cm]{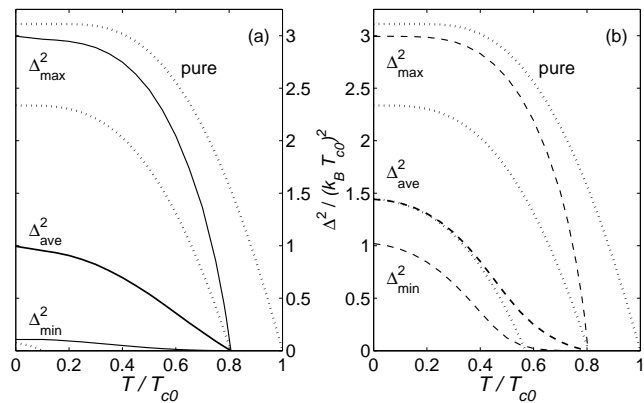}
\caption{Temperature dependence of the squared order parameter in the B 
phase. The solid lines give the minimum ($\Delta_{\rm min}^2$),
the maximum ($\Delta_{\rm max}^2$), and the average value 
[$\Delta_{\rm ave}^2$, Eq. (\ref{e.deltaave})]. They are calculated in
IISM for void impurity profile  [Eq. (\ref{e.idens1})] with
$R = 2\ell_{\rm ave}$. The left panel uses the steep profile ($j = 8$) and
$\xi_0/\ell_{\rm ave} = 0.275$ and the right the gentle profile ($j = 2$) and
$\xi_0/\ell_{\rm ave} = 0.16$. Both cases have the same
transition temperature ($T_{\rm c} = 0.81 T_{\rm c0}$). The three dashed
lines in both panels are calculated with the HSM. From larger to smaller they
correspond to the pure limit ($\xi_0/\ell = 0$), the HSM with the same $T_{\rm
c}$ ($\xi_0/\ell = 0.076$) and HSM with the same average scattering rate as
in the IISM ($\xi_0/\ell=0.275$ in the left panel and 0.16 in the
right panel). }       
\label{f.gapesim2}
\end{center}
\end{figure}

Let us now turn to the amplitude $\Delta$ of the order parameter. In the HSM
$\Delta/k_{\rm B}T_{\rm c}$ depends mostly on the relative temperature 
$T/T_{\rm c}$ but also on the scattering rate $\xi_0/\ell$, the phase shifts
$\delta^{(l)}$, and the phase (A or B).\cite{PhysicaB} The dotted lines in
Fig.\ \ref{f.gapesim2} represent four different values of $\xi_0/\ell$
for B-phase with $\sin^2\delta^{(0)} =\frac{1}{2}$.   

In the IISM the parameter $\xi_0/\ell$ is replaced by $\xi_0/\ell_{\rm
ave}$ and $n(r)$. The order parameter depends
on the location ${\bf r}$ (\ref{ordparBvszero}). In order to describe
its temperature dependence we use parameters
$\Delta_{\rm max}$ and $\Delta_{\rm min}$, and $\Delta_{\rm
ave}$ (\ref{e.deltaave}). The temperature dependence of the three
characteristic numbers is shown by solid lines in Fig.\ \ref{f.gapesim2}.
We see that there can be strong variation since $\Delta_{\rm min}^2$
and $\Delta_{\rm max}^2$ are quite different. The average
$\Delta_{\rm ave}^2$ is considerably below the HSM curve that gives
the same $T_{\rm c}$. Therefore $\Delta_{\rm ave}^2$ is more
suppressed than the critical temperature $T_{\rm c}$. 

The temperature dependence of $\Delta_{\rm ave}^2$ varies. Let us first
consider the case  in the right-hand panel of Fig.\
\ref{f.gapesim2}, where the shape of $\Delta_{\rm ave}^2(T)$
is concave near $T_{\rm c}$. This is typical for a small inhomogeneity 
amplitude. In our case it also means that the
average scattering is small since the minimum scattering vanishes. We see
that at low temperatures $\Delta_{\rm ave}^2$ agrees very well with the HSM
curve that is calculated with the same {\em average} scattering rate. At
higher temperatures $\Delta_{\rm ave}^2$ deviates from this since the
true $T_{\rm c}$ is higher than the one given by the average scattering rate.

In the case of the left hand panel of Fig.\
\ref{f.gapesim2}, the temperature dependence of $\Delta_{\rm ave}^2$
is quite linear near $T_{\rm c}$. In fact, the linear range is wider
(relative to $T_{\rm c}$) than in the pure case or HSM. This happens
when the amplitude of the inhomogeneity is large. Large
inhomogeneity means large average scattering, thus the HSM results based on
the average scattering rate  are more suppressed than in the case of
concave $\Delta_{\rm ave}^2(T)$. In the particular case of Fig.\
\ref{f.gapesim2}(a), the HSM result is quite small
since  $\xi_0/\ell=0.275$  is near the critical value of complete
suppression. In this case
$\Delta_{\rm ave}^2$ is nowhere nearly approximated by the HSM result.
However, the linear $\Delta_{\rm ave}^2(T)$ can also appear in cases where
the HSM with average scattering rate is not that suppressed, and provides a
good approximation for $\Delta_{\rm ave}^2$ at low temperatures. This takes
place at small $R/\xi_0$, where the proximity coupling between different
regions tends to average out the inhomogeneity. In this limit the range of
scattering rates $\xi_0/\ell_{\rm
ave}$ where $\Delta_{\rm ave}^2(T)$ is concave near $T_{\rm c}$ seems to
vanish.

The concave shape of $\Delta_{\rm ave}^2(T)$ in Fig.\
\ref{f.gapesim2}(b) can be understood so that the different
regions of an inhomogeneous sample have transitions more or less
independently of each other: At
$T_{\rm c}$ only $\Delta(r=0)=\Delta_{\rm max}$ starts to grow but
$\Delta_{a}(r=R)=\Delta_{\rm min}$ and $\Delta_{r}(R)$ (not shown)
remain negligible until they start to grow at a lower
temperature. In spite of the inhomogeneity, the onset of superfluidity
indicated by $\Delta_{\rm max}$ is very sharp giving a well
defined $T_{\rm c}$.

The superfluid density $\rho_{\rm s}$ is conveniently expressed 
relative to the density of the liquid, $\rho_{\rm s}/\rho_{\rm tot}$. In
addition to parameters discussed above, this depends on the
Fermi-liquid parameter $F_1^{s}$ in both the HSM and IISM. (In the
IISM also other parameters could contribute, but they are neglected
here.)  Since $F_1^{s}\sim 10$ is quite big, it has a strong effect on
the results. The HSM results for $\tensor
\rho_s$ in both the A and B phases have been calculated in Refs.
\onlinecite{PhysicaB} and \onlinecite{higashitani}. 

\begin{figure}[tb]
\begin{center}
\includegraphics[width=8.5cm]{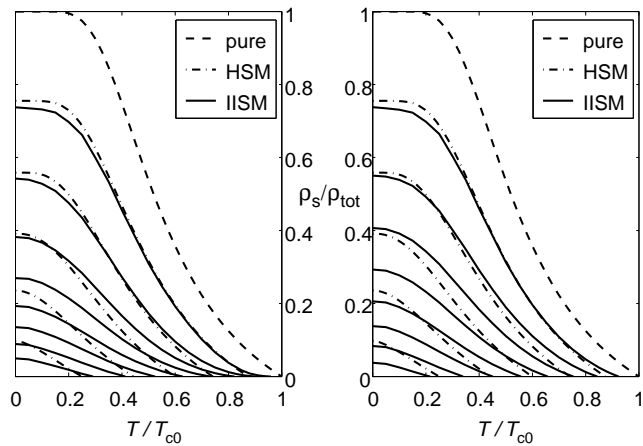}
\caption{Superfluid density in the B phase at $F_1^s = 10$
(corresponding to $^3$He at 1.56 MPa). The uppermost dashed line is
for pure superfluid. The dash-dotted lines are for the HSM with
$\xi_0/\ell$ = 0.05, 0.10, 0.15, 0.20, and 0.25 in order of decreasing
$T_{\rm c}$. The solid lines are the IISM results corresponding to
$\xi_0/\ell_{\rm ave}$ = 0.05, 0.10, ..., 0.40 and $R = \ell_{\rm ave}$.
On the left we have used the steep void scattering profile [Eq.\
(\ref{e.idens1}) with
$j=8$] and on the right the cluster profile (\ref{e.idens2})
with $j=3$ and $b=0.1$. }       
\label{f.rhosesim}
\end{center}
\end{figure}

In the IISM we calculate $\rho_{\rm s}$ for the B phase. Some curves are
plotted in Fig.~\ref{f.rhosesim}.
As above, we compare the IISM with the HSM calculated with
the same average scattering rate $\xi_0/\ell_{\rm ave}$. The results
of both models are close to each other at low
temperatures when the scattering rate is small. At higher temperatures
$\rho_{\rm s}$ in IISM is larger because of its higher
$T_{\rm c}$. Thus $\rho_{\rm s}$ is relatively more suppressed than $T_{\rm
c}$. At larger scattering rates, $\rho_{\rm s}$ in the HSM becomes smaller
and is completely suppressed at $\xi_0/\ell=0.28$, while
$\rho_{\rm s}$ in the IISM stays finite much beyond that. 
All these results are qualitatively similar as discussed above for 
$\Delta_{\rm ave}^2$. The discussion of
concave or linear temperature dependence in the case of $\Delta_{\rm
ave}^2$ cannot directly be applied to $\rho_{\rm s}$ since this behavior
is largely masked by the effect of
$F_1^{s}$. The superfluid density $\rho_{\rm s}$ is also different from
$\Delta_{\rm ave}^2$ because of current conservation, which forces
quite different current pattern in the void and cluster
impurity profiles. 

In order to reduce the effect of $F_1^s$, we define a bare superfluid
density $\rho_{\rm s}^{\rm b}$ by  
\begin{eqnarray}
\frac{\rho_{\rm s}^{\rm b}}{\rho_{\rm tot}} = 
\frac{ \left(1+\tfrac{1}{3}F_1^s\right) ({\rho_{\rm s}}/{\rho_{\rm tot}}) }
{ 1+\tfrac{1}{3}F_1^s ({\rho_{\rm s}}/{\rho_{\rm tot}}) } .
\label{e.rhobaradef}
\end{eqnarray}
In the case of the HSM this is independent of $F_1^s$ and in the case of 
the IISM the dependence is small, only a few percent over the whole pressure
range $0\ldots 3.4$ MPa. We calculate $\rho_{\rm s}^{\rm b}$ at an
intermediate pressure of 1.56 MPa where
$F_1^s = 10$. 

\begin{figure}[tb]
\begin{center}
\includegraphics[width=8.5cm]{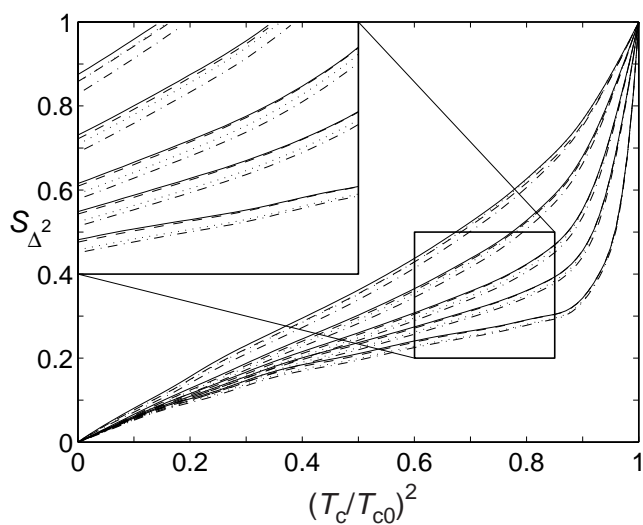}
\caption{Suppression factor for the squared order parameter in different
cases: 
B phase $\Delta_{\rm ave}^2$ (solid lines),  
B phase $\Delta_{\rm NMR}^2$ (dotted lines),  
A phase $\Delta_{\rm ave}^2$ (dashed lines), and 
A phase $\Delta_{\rm NMR}^2$ (dash-dotted lines). The five sets of curves
correspond to reduced  temperatures $t=0.2$, 0.5,
0.7, 0.8 and 0.9, from top to bottom.
Other parameters are steep void profile [Eq.\
(\ref{e.idens1}) with $j=8$] and $R = \ell_{\rm ave}$.
The inset shows double magnification of the part indicated by the
rectangle.}
\label{f.suppdb2da2}
\end{center}
\end{figure}

The order parameter and superfluid density can be expressed compactly by 
defining suppression factors 
\begin{eqnarray}
S_{\Delta^2}(t) &=& \frac{\Delta^2(t T_{\rm c})}
{\Delta_{\rm 0}^2(t T_{\rm c0})} \nonumber \\
S_{\rho_{\rm s}^{\rm b}}(t) &=& \frac{\rho_{\rm s}^{\rm b}(t T_{\rm c})}
{\rho_{\rm s0}^{\rm b}(t T_{\rm c0})} .
\label{e.suppfactordef}
\end{eqnarray}
Here $\rho_{\rm s0}^{\rm b}$ is the bare superfluid density and
$\Delta_{\rm 0}^2$ the order parameter, both in pure superfluid. The
parameter $t$ is the temperature relative to the transition
temperature. Additionally, we wish to eliminate the parameter
$\ell_{\rm ave}$, which is not directly measurable. This is achieved
in plotting the suppression factors as a function of 
$(T_{\rm c}/T_{\rm c0})^2$. 
One such plot is shown in Fig.~\ref{f.suppdb2da2}. 

The important conclusion from Fig.~\ref{f.suppdb2da2} is that the
suppression factors for the A and B phases are almost identical,
although there is a clear difference in the  values of $\Delta_{\rm
ave}$. This generalizes the result found previously within
the HSM.\cite{PhysicaB} Also, the difference in $S_{\Delta^2}$ between
$\Delta_{\rm ave}$ and $\Delta_{\rm NMR}$ is small. The differences
increase with increasing inhomogeneity, so that the suppression
factors for $\Delta_{\rm ave}^2$ and $\Delta_{\rm NMR}^2$ differ by
$\approx 5$\% in our extreme case $R=2\ell_{\rm ave}$. All these
differences are rather small and, in order to simplify the plots, we
present below $S_{\Delta^2}$ only for the B phase $\Delta_{\rm ave}$.

\begin{figure}[tb]
\begin{center}
\includegraphics[width=8.5cm]{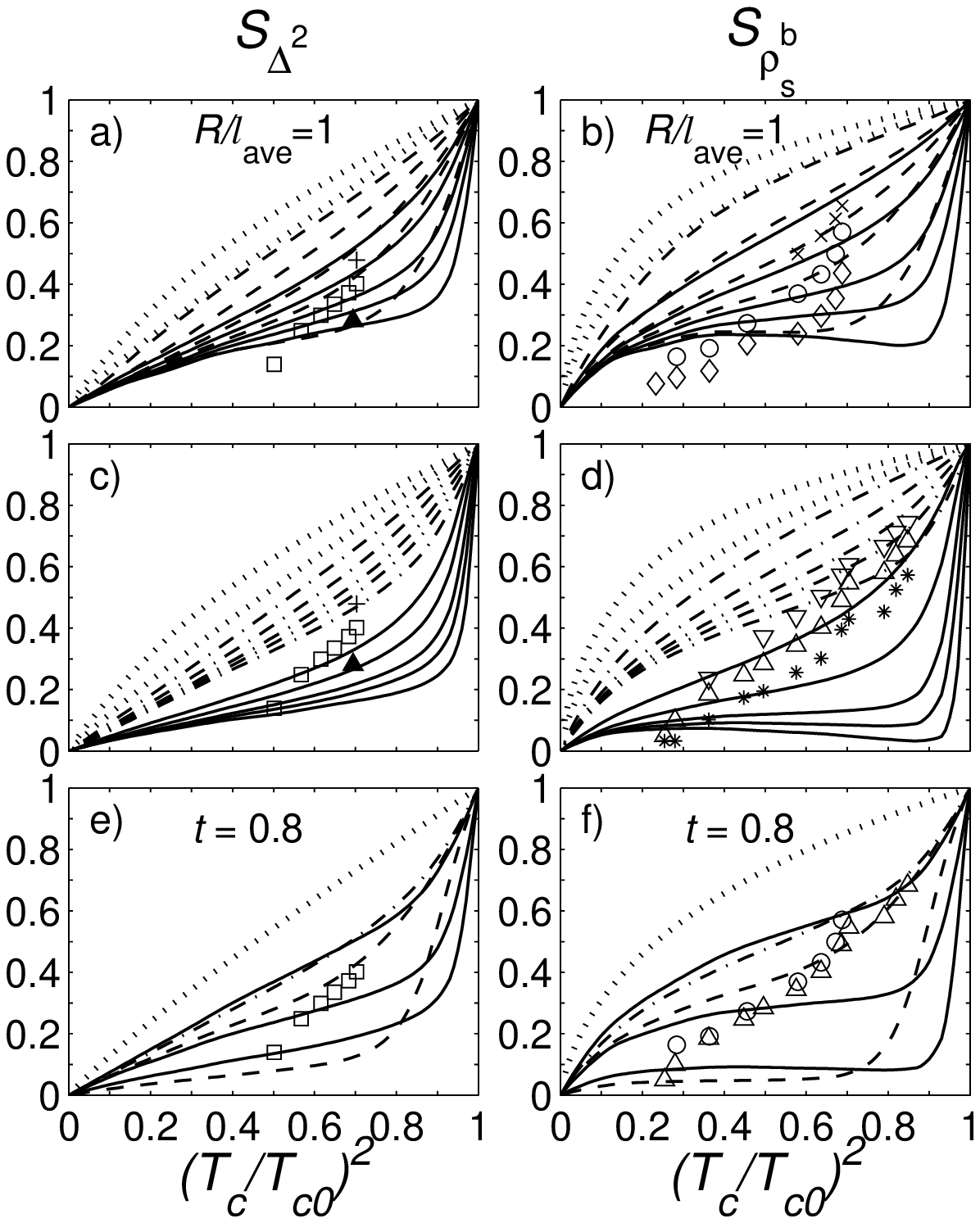}
\caption{Suppression factors (\protect\ref{e.suppfactordef}) for the squared
order parameter [Eq. (\protect\ref{e.deltaave})] (left) and for the bare
superfluid density (\protect\ref{e.rhobaradef}) (right). The solid and
dashed lines are for void impurity profile (\protect\ref{e.idens1}) with steep
($j=8$) and gentle ($j=2$) slopes, respectively.  The dash-dotted lines are
for the cluster profile (\protect\ref{e.idens2}) with
$j=3$ and $b=0.1$. The dotted lines are calculate using the HSM. For subplots
$a$--$d$ the temperatures from top to bottom are $t$ =  0.2, 0.5, 0.7, 0.8, and
0.9 for the IISM. For HSM the upper line is for
$t\rightarrow 0$ and the lower one for $t\rightarrow 1$. The subplots $e$
and $f$ compare the different profiles at $t = 0.8$. The radius of the
unit cell is mainly $R = \ell_{\rm ave}$ except for subplots $c$ and $d$
where the solid lines are for $R = 2\ell_{\rm ave}$ and for subplots $e$
and $f$ where the uppermost solid line is for $R = 0.5\ell_{\rm ave}$ and
the lowest solid and dashed lines are for $R = 2\ell_{\rm
ave}$. Experimental points for the order parameter are from
Refs.~\protect\onlinecite{Sprague95} and
\protect\onlinecite{Sprague96}  with  $t=0.9$ 
($\protect\Box$) and $t = 0.5$ ($+$) and \protect\onlinecite{Barkerprl}
($\protect\blacktriangle$, $t = 0.65$), see discussion in the main text. The
measured superfluid densities are from Manchester
(Ref. \onlinecite{Hook}) with 
$t = 0.5$ $(\protect\times)$, $t = 0.8$ $(\protect\circ)$, and $t =
0.9$ $(\protect\Diamond)$ and from Cornell sample A
(Ref. \onlinecite{Porto99}),
with $t = 0.5$ $(\protect\triangledown)$, $t = 0.8$
$(\protect\vartriangle)$, and $t = 0.9$ $(*)$.}   
\label{f.supp}
\end{center}
\end{figure}

Suppression factors for both $\Delta^2$ and $\rho_{\rm s}^{\rm b}$ as
functions of $(T_{\rm c}/T_{\rm c0})^2$ are plotted in Fig.~\ref{f.supp}. 
The HSM results are shown for comparison. They are all above the
diagonal. It is clearly visible that the effect of inhomogeneity is to
bend these curves down. This means that $\rho_{\rm s}^{\rm b}$ and
$\Delta^2$ are  more strongly suppressed than $T_{\rm c}$. With a
steep profile and a large $R/\ell_{\rm ave}$ one obtains a strong
suppression and a wide flat region in the suppression curve. 
Superfluid density is clearly more sensitive to the inhomogeneity than
the order parameter.  

\section{Comparison with experiments}\label{s.exp}

\begin{figure}[tb]
\begin{center}
\includegraphics[width=8.5cm]{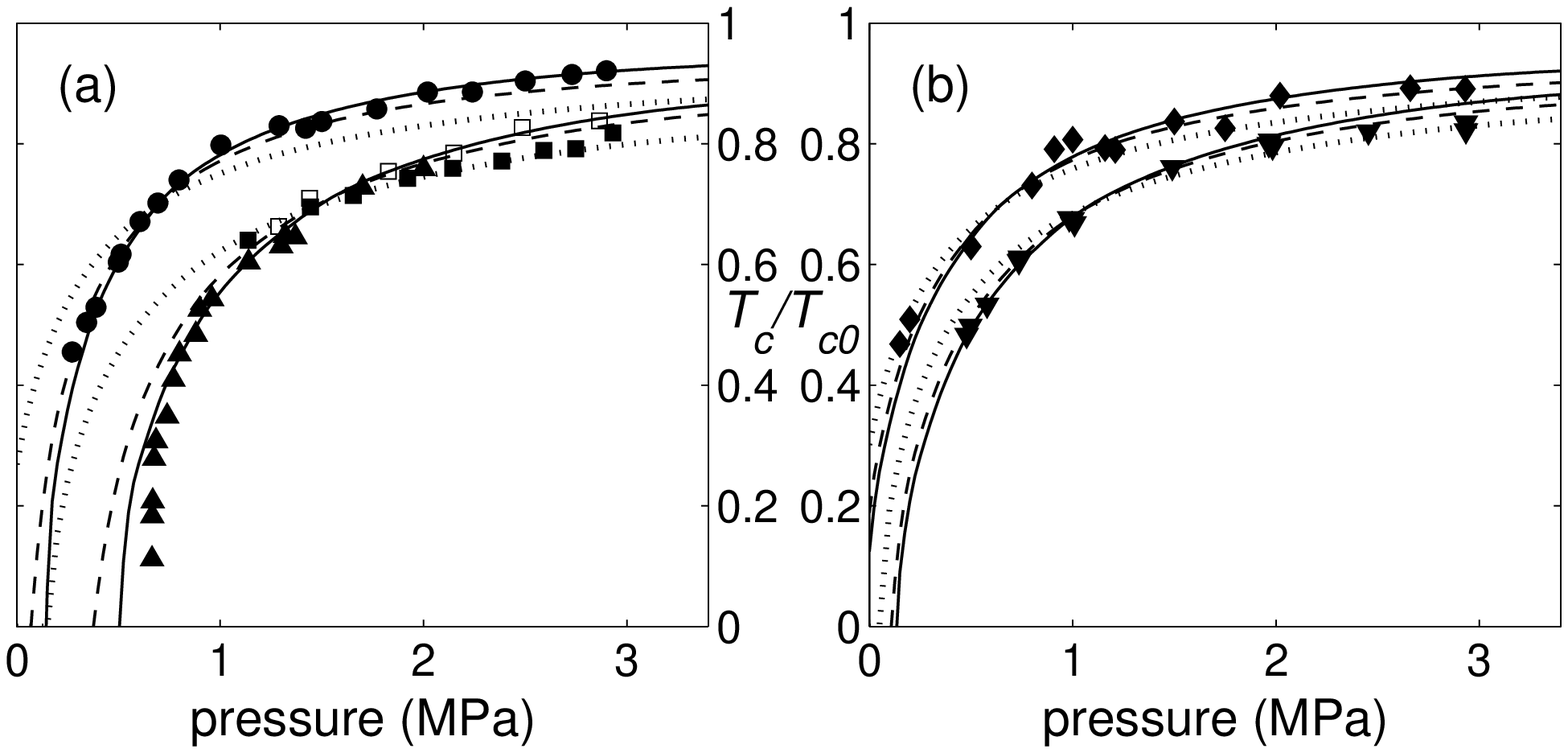}
\caption{Comparison of the calculated transition temperatures with
experiments. The solid  and dashed lines  are calculated using the void
impurity profile  (\ref{e.idens1}) with steep
($j  = 8$) and gentle ($j = 2$) slopes, respectively. The dotted lines are
from the HSM. In panel (a) the experimental points are from
Refs.~\protect\onlinecite{Porto95} ($\protect\bullet$, Cornell sample A),
\protect\onlinecite{Porto99} ($\protect\blacksquare$), 
\protect\onlinecite{Matsumoto} ($\protect\blacktriangle$, Cornell
sample C), and \protect\onlinecite{Sprague95} and
\protect\onlinecite{Sprague96} ($\protect\square$). 
Here $\ell$ = 213 (314) nm for the HSM,
$R = 2\ell_{\rm ave}$ = 140 (206) nm for $j = 8$ and $R = 2\ell_{\rm
ave}$ = 234 (342) nm for $j = 2$. The values in the brackets denote the
upper curves. In panel (b) the measured points are from
Ref.~\protect\onlinecite{Hook}
($\protect\blacklozenge,\protect\blacktriangledown$) and we 
have  used $\ell$ = 251 (324) nm for the HSM, $R = \ell_{\rm ave}$ =
132 (170) nm for $j = 8$, and $R = \ell_{\rm ave}$ = 195 (252) nm for $j = 2$. 
All experimental points are for aerogel with nominally 98\% open volume
except $\protect\blacklozenge$, which are for 99\%. Small-angle x-ray
scattering gives the aerogel correlation lengths $\xi_a= 84$ and
130 nm for samples denoted by $\protect\blacktriangle$ and
$\protect\bullet$, respectively (Ref. \onlinecite{Porto99}).} 
\label{f.experimtc}
\end{center}
\end{figure}

A comparison of the calculated transition temperatures with experimental
results is given in Fig.~\ref{f.experimtc}. 
The pressure-independent $\ell_{\rm ave}$ is determined  so
that the theoretical curves and the experimental data intersect at the point
where $T_{\rm c}/T_{\rm c0} = 0.7$.  For the samples measured in Manchester
[Fig.~\ref{f.experimtc}b] the HSM gives a fairly good fit but a better
correspondence is obtained with IISM by using $R =
\ell_{\rm ave}$. For samples measured in Cornell and Northwestern 
[Fig.~\ref{f.experimtc}a] the closest curve is the IISM
with $R = 2\ell_{\rm ave}$ and $j = 8$ in  Eq.~(\ref{e.idens1}). 
This represents roughly the best fit obtained by the IISM.

Particularly interesting are the samples A and C measured in Cornell 
[Fig.~\ref{f.experimtc}(a)], which also have been studied using small-angle
x-ray scattering. The values measured for the aerogel correlation
scale $\xi_{\rm a}$ are 130 and 84 nm.\cite{Porto99} These are 
slightly smaller but on the same order of magnitude as our closest curve
values $R = 210$ and 140 nm, respectively. This supports the view that
the better fit obtained by the IISM as compared to the HSM is not only
due to more fitting parameters, but a more realistic modelling of the
structure of the aerogel.   

Some experimental points for both the order parameter and superfluid
density are shown in Fig.~\ref{f.supp}. Each marker ($\triangledown$,
$\vartriangle$, $*$ etc.) identifies a data set that corresponds
to one aerogel sample at a fixed reduced temperature
$t$. The different points in each data set are obtained from
measurements at different pressures.  The superfluid density is
measured by torsional oscillator and the interpretation of the
experiments is rather straightforward. The order parameter is measured
by NMR. This is possible because the frequency shift depends on the
dipole-dipole energy, Eqs.\ (\ref{e.ddb}) or (\ref{e.dda}). For
interpretation one has to know whether the phase is A or B. The
Stanford point ($\blacktriangle$) is based on seeing similar
suppression in both phases.\cite{Barkerprl} The Northwestern data is
drawn here by assuming A phase and $\hat{\bf d}$ and
$\hat{\bf l}$ perpendicular to magnetic field, as they are in bulk
$^3$He-A.\cite{Sprague95,Sprague96} However, it is not clear if this original
assumption is correct.\cite{halprev} In case of the B phase the
analysis would be more complicated because it would require an analysis of
the texture.\cite{Alles,Barkerprl,btex}

One can notice that the experimental data for the order
parameter and superfluid density are qualitatively similar. They all fall
clearly below the diagonal, and appear to extrapolate to $S=0$ already at
a finite $T_{\rm c}/T_{\rm c0}$. This is in clear disagreement with the HSM
results that are all above the diagonal. 
It can be seen that much better agreement is achieved with the IISM.
Ideally each experimental data
set should fall on some of the theoretical lines. As one can see, this is
not quite the case in the IISM. A reasonable over-all fit to suppression factors
and critical temperatures is achieved with the  void impurity profile
(\ref{e.idens1}) with $j =  2$ and $R\approx 1.5\ell_{\rm ave}$, but the
choice depends on the properties one wishes to emphasize. The agreement could
be improved by allowing for a pressure-dependent mean free path $\ell$.
However, the most obvious reason for the remaining differences is that the
scattering profile in the IISM has only one length scale whereas real aerogel
has a wide distribution of length scales.

\section{Conclusions}\label{s.con}

We have presented the isotropic inhomogeneous scattering model. We claim
it is the simplest model of inhomogeneous scattering that is consistent
with large-scale isotropy. Unfortunately, the computations needed are
much more demanding than in the homogeneous scattering model. The model
itself is independent of the pairing symmetry and thus could be used to
study the effect of inhomogeneous impurity distributions in superconductors. 

When applied to $^3$He in aerogel, the IISM gives better agreement with
experiments than the HSM. We emphasize that this is not solely due to the
IISM having more free parameters than the HSM. On the contrary, the
inhomogeneity of aerogel is the most natural if not the only way to explain
the differences between measurements and the HSM. The fitted  parameters of
IISM are in reasonable agreement with measurements and simulations on the
structure of aerogel. The fit given by IISM is not perfect, though. The main
problem with the IISM is that it contains only one length scale whereas real
aerogel must have voids of various sizes. 

\acknowledgments
We wish to thank V. Dmitriev, W. Halperin, J. Hook, G. Kharadze, D. Osheroff,
J. Parpia, D. Rainer, J. Sauls and J. Viljas for useful discussions. 
We also thank the Academy of Finland for financial support and CSC for
providing computer resources.   

\appendix
\section{}

Here we present the relevant equations of the quasiclassical theory and
discuss their numerical solution.   The central quantity is the
quasiclassical propagator $\breve{g}$.\cite{Serene} It is a $4\times
4$ matrix, whose components can be represented using the Pauli spin
matrices $\spin{\sigma}_i$ as 
\begin{eqnarray}
{\nambu g}=\left(
\begin{array}{cc}
g+{\bf g}\cdot\spin{\boldsymbol{\sigma}} & (f+{\bf f} \cdot
\spin{\boldsymbol{\sigma}})\imagu\spin{\sigma}_2  \\
\imagu\spin{\sigma}_2({\tilde f}+{\tilde{\bf f}}\cdot
\spin{\boldsymbol{\sigma}}) & {\tilde g} -
\spin{\sigma}_2{\tilde {\bf g}}
\cdot \spin{\boldsymbol{\sigma}}
\,\spin{\sigma}_2
\end{array}
\right).
\label{e.propadef1}
\end{eqnarray} In equilibrium the propagator 
$\breve{g}(\hat{\bf k},{\bf r},\epsilon_m)$ depends on the direction of
momentum $\hat{\bf k}$, on the location ${\bf r}$, and on the Matsubara
energy $\epsilon_m=\pi k_{\rm B} T(2m-1)$, where $m$ is an
integer. The propagator is determined by the Eilenberger equations,
\begin{eqnarray}
[{\rm i}\epsilon_m\breve{\tau}_3-\breve{\sigma},\breve{g}]+{\rm i}\hbar
v_{\rm F} \hat{\bf k} \cdot \nabla_{\bf r} \breve{g} &=& 0 ,
\label{e.eil1}  \\ 
\breve{g}\breve{g} &=& -1.
\label{e.eilenb}
\end{eqnarray}
Here $[A,B] = AB - BA$ denotes a commutator and $\breve{\tau}_i$
denote Pauli matrices in the Nambu space; $\breve{\tau}_i=\spin{\sigma}_i
\otimes 1$. The self energy matrix $\breve{\sigma} = \breve{\sigma}_{\rm mf}
+\breve{\sigma}_{\rm imp}$ consists of mean field and impurity
contributions. For spin-triplet pairing the former part has the form    
\begin{equation}
\breve{\sigma}_{\rm mf}(\hat{\bf k},{\bf r})=\left( 
\begin{array}{cc}
\nu(\hat{\bf k},{\bf r}) + \boldsymbol{\nu}(\hat{\bf k},{\bf r}) \cdot
\spin{\boldsymbol{\sigma}} & 
\boldsymbol{\Delta}(\hat{\bf k},{\bf r})\cdot \spin{\boldsymbol{\sigma}}
{\rm i} \spin{\sigma}_2 \\ 
{\rm i}\spin{\sigma}_2 \boldsymbol{\Delta}^*(\hat{\bf k},{\bf r})\cdot
\spin{\boldsymbol{\sigma}} &  {\nu}(-\hat{\bf
k},{\bf r}) - \spin{\sigma}_2
{\boldsymbol{\nu}}(-\hat{\bf k},{\bf r}) \cdot
\spin{\boldsymbol{\sigma}}\, \spin{\sigma}_2   
\end{array} 
\right).
\label{e.purese}
\end{equation}
The real-valued functions $\nu$ and $\boldsymbol{\nu}$ depend on the
Fermi liquid parameters $F_l^{s,a}$. 
Omitting $F_3^s$, $F_1^a$ and higher coefficients we have
$\boldsymbol{\nu}=0$ and 
\begin{eqnarray}
\nu(\hat{\bf k},{\bf r}) &=& \frac{\pi k_{\rm B} T F_1^{s} }
{ 1+\frac{1}{3}F_1^{s}}\sum_{\epsilon_m}
\left\langle(\hat{\bf k} \cdot \hat{\bf k}') 
g(\hat{\bf k}',{\bf r},\epsilon_m)\right\rangle_{\unit{\bf k}'},
\end{eqnarray}
where $\langle\cdots\rangle_{\unit{\bf k}}$ denotes angular average. The
off-diagonal part of $\breve{\sigma}_{\rm mf}$ defines the order parameter,
which for $p$-wave pairing is 
$\boldsymbol{\Delta}=\tensor A\cdot \hat{\bf k}$. It is determined by the
weak-coupling self-consistency equation 
\begin{eqnarray}
\boldsymbol{\Delta}(\hat{\bf k},{\bf r}) \ln{\frac{T}{T_{\rm c0}}} + \pi
k_{\rm B} T
\sum_{\epsilon_m}  \left\lbrack \frac{\boldsymbol{\Delta}(\hat{\bf k},{\bf
r}) }{\vert
\epsilon_m \vert}  - 3 \left\langle
(\hat{\bf k}\cdot\hat{\bf k}') {\bf f} (\hat{\bf k}',{\bf
r},\epsilon_m)\right\rangle_{\unit{\bf k}'}\right\rbrack = 0. 
\label{e.gapequ} 
\end{eqnarray}
The impurity self energy $\breve{\sigma}_{\rm imp}$ is related to the 
forward-scattering $t$ matrix via 
\begin{equation}
\breve{\sigma}_{\rm imp}(\hat{\bf k},{\bf r},\epsilon_m) = n({\bf r})
\breve{t}(\hat{\bf k},\hat{\bf k},{\bf r},\epsilon_m),
\end{equation}
where $n({\bf r})$ is the impurity density. The usual
derivation of this formula assumes a uniform impurity density
$n ={\rm const}$.\cite{AGD} However, it is valid for an arbitrary
function $n({\bf r})$ as long as it allows the impurity positions to
be uncertain on the scale of the Fermi wave length
$\lambda_{\rm F}$. We assume that only $s$-wave scattering is
important. This allows us to write the $t$-matrix equation directly for
the impurity self-energy
\begin{equation}
\breve{\sigma}_{\rm imp}({\bf r},\epsilon_m) = n({\bf r}) v \breve{1} 
+\pi N_{\rm F} v \left\langle \breve{g}(\hat{\bf k},{\bf
r},\epsilon_m)\right\rangle_{\unit{\bf k}}
\breve{\sigma}_{\rm imp}({\bf r},\epsilon_m).  
\label{e.tequ}
\end{equation}
Here $2N_{\rm F}\! = \!m^*k_{\rm F}/\pi^2\hbar^2$ is the total
density of states at the Fermi surface. The effective mass $m^*$ is
related to the atomic mass $m_3$ by $m^*=( 1+\frac{1}{3}F_1^{s})m_3$. The
scattering potential $v$ is related to the scattering phase shift
$\delta^{(0)}$ by $v=-(1/\pi N_{\rm F}) \tan{\delta^{(0)}}$. 

In the calculation we use the following symmetry of the propagator and
self energy
\begin{equation}
\lbrack \breve{u}(\hat{\bf k},{\bf r},\epsilon_m) \rbrack ^{T} =
-\breve{\tau}_2 \breve{u} (\hat{\bf k},{\bf r},\epsilon_m) \breve{\tau}_2 .
\label{e.symmetry2}
\end{equation}
This limits the nonzero components of $\breve{g}$ to 10, and allows one to
solve the Eilenberger equation with the ``multiplication trick'' of Ref.\
\onlinecite{Zhang} using only five components. In the current-free case
all equations are immediately compatible with the
symmetry in Eq.\ (\ref{e.symmetry2}). In the case of nonzero current a
possible problem could arise from the fact that in iterating the impurity
equation (\ref{e.tequ}) the product $\langle \breve{g}\rangle_{\unit{\bf
k}}\langle \breve{g}\rangle_{\unit{\bf k}}$ might not be proportional to a unit matrix
$\breve{1}$. The inconvenient terms always appear with odd powers of $v$.
Considering that the phase shifts in aerogel are random, such terms average
out and we can use the symmetry of Eq.\  (\ref{e.symmetry2}) also for
nonzero current.  

By taking into account all the symmetries the impurity self energy in
the IISM has the form
\begin{eqnarray}   
\breve{\sigma}_{\rm imp}({\bf r},\epsilon_m) = 
\left( \begin{array}{cc} 
{\rm i}\delta+{\rm i}\boldsymbol{\delta} \cdot \spin{\boldsymbol{\sigma}} &
{\bf R} \cdot \spin{\boldsymbol{\sigma}} {\rm i}\spin{\sigma}_2 \\
-{\rm i}\spin{\sigma}_2 {\bf R}^* \cdot \spin{\boldsymbol{\sigma}} & -{\rm
i}\delta + \spin{\sigma}_2 {\rm i} \boldsymbol{\delta} \cdot
\spin{\boldsymbol{\sigma}}\,
\spin{\sigma}_2 \end{array} \right) + w \breve{1},
\label{e.rhoform}
\end{eqnarray}
where ${\bf R}$, $\boldsymbol{\delta}$, and $\delta$ can be expressed
analytically as functions of the propagator $\langle
\breve{g}\rangle_{\unit{\bf k}}$.\cite{diplomit}
 The  term $w$ is of no interest
since it drops out in the Eilenberger  equation (\ref{e.eil1}).  In
cylindrical coordinates ($\rho$, $\phi$, $z$) the only
$\phi$ dependence appears in the vector ${\bf R}$  in the $A$ phase, where
${\bf R} = R_0(\rho,z)e^{{\rm i}
\phi}\hat{\bf z}$ and $R_0$ is real. The terms $\delta$ and
$\boldsymbol{\delta}$ are always real. In the A phase
$\boldsymbol{\delta} = 0$ and in the $B$ phase $\boldsymbol{\delta}
=\delta_\phi \hat{\boldsymbol{\phi}}$ and ${\bf R} = R_\rho
\hat{\boldsymbol{\rho}}+R_z\hat{\bf z}$.    

From the propagator ${\nambu g}$ (\ref{e.propadef1}) one can calculate
the supercurrent density ${\bf j}_{\rm s}$ using     
\begin{equation}
{\bf j}_{\rm s}({\bf r}) = 2 \pi m_3 v_{\rm F} N_{\rm F} k_{\rm B} T
\sum_{\epsilon_m}\left\langle {\hat {\bf k}} 
g({\hat {\bf k}},{\bf r},\epsilon_m)\right\rangle_{\unit{\bf k}}. 
\label{e.masscurrent}
\end{equation}
The superfluid density $\rho_s$ is obtained by calculating the
averaged current at small values of ${\bf v}_{\rm s}$.

The first step in the calculation is to give initial values for the fields
$\tensor{A}$, $\breve{\sigma}_{\rm imp}$, and $\nu$. Normally we used values
obtained from the HSM. The iteration of these fields was started by
transforming the data from the cylindrical/spherical coordinates to
``trajectory coordinates.'' These are Cartesian coordinates with one
coordinate, $u$, along the trajectory and two others ($b$ and $t$) specifying
the position of the trajectory in the unit cell. Next  the Eilenberger
equation was solved along the trajectories.  This was done by first
calculating an unphysical exponentially growing propagator along each
trajectory using the fourth order Runge-Kutta method. The end values at the
cell boundary were saved to be used as starting values for next step in
iteration. An exponentially decreasing solution was deduced using
symmetries. The bounded physical propagator was obtained as a commutator of
these two unphysical solutions as explained in Ref.\
\onlinecite{Zhang}. The solution of the Eilenberger equation was
repeated for all Matsubara frequencies $\vert\epm\vert \le
\epsilon_{N}$. This was repeated for all trajectories. In the
general case it represents loops for  coordinates $b$, 
$t$ and the angle $\alpha$  between the trajectory direction and
the $z$-axis. For the B phase without a current only the impact parameter $b$
with respect to the center of the sphere was actually needed because of
symmetry.

In the next step better estimates for $\tensor{A}$,
$\breve{\sigma}_{\rm imp}$, and $\nu$ were calculated. For that the
propagator was converted from trajectory coordinates back to
cylindrical/spherical coordinates and the required angular averages were
calculated. The contribution from higher Matsubara frequencies
$\vert\epm\vert > \epsilon_{N}$ was approximated by a Ginzburg-Landau
form \cite{ErkkiGL} with Matsubara sums evaluated
using the Euler-MacLaurin formula. 
Using the updated values of the fields, the process was started from the
beginning.  The boundary condition (\ref{e.boundary}) was used to calculate
the value of the exponential propagator at the initial point of the
trajectory from the value at the final point stored on the previous round.
This loop was repeated until the fields converged. 

In the numerical algorithm the number of discretized points in
cylindrical coordinates $\rho_i$ and $z_j$ was approximately 40 for both 
in the range from $0$ to $R$. The discretization step
in the Cartesian trajectory coordinates was approximately the same. Simple
interpolation formulas were used for the transformations between coordinate
systems. The angle $\alpha$ was typically discretized by eight Gaussian points
in the range $0<\alpha<\pi/2$.  The number of positive Matsubara frequencies
$\epsilon_m$ used was typically less than 20 for temperatures above $0.1
T_{\rm c0}$. We made also more accurate test calculations. The qualitative
behavior remains the same, but there are inaccuracies on the order of
2\% in the results presented here.

\end{document}